\documentstyle[twoside,fleqn,espcrc2,psfig]{article}

   \newcommand{\Z}{{\cal Z}}
   \newcommand{\be}{\begin{equation}}
   \newcommand{\ee}{\end{equation}}
   
\def\Z{{\cal Z}}
\def\t{\tilde}

\newcommand{\AmS}{{\protect\the\textfont2
  A\kern-.1667em\lower.5ex\hbox{M}\kern-.125emS}}

\title{No-go theorem on spontaneous parity breaking revisited}

\author{V. Azcoiti\address{Departamento de F\'isica Te\'orica,
	Facultad de Ciencias,Universidad de Zaragoza,50009 Zaragoza,Spain},
	A. Galante$^a$\thanks{Talk presented by A. Galante}}
 
\begin{document}
\pagestyle{empty}
\begin{abstract}
An essential assumption in the Vafa and Witten's theorem on P and CT
realization in vector-like theories concerns the existence of a free 
energy density in Euclidean space in the presence of any external 
hermitian symmetry breaking source. We show how this requires the 
previous assumption that the symmetry is realized in the vacuum. 
Even if Vafa and Witten's conjecture is plausible, actually a theorem 
is still lacking. 
\end{abstract}

\maketitle
%\section*{Introduction}

A few years ago Vafa and Witten gave an argument against spontaneous
breaking of parity in vector-like parity-conserving theories as QCD 
\cite{WITTEN}. 
The main point in their proof was the crucial observation that any 
arbitrary hermitian local order parameter $X$ constructed from Bose fields 
should be proportional to an odd power of the four indices antisymmetric 
tensor $\epsilon^{\mu\nu\rho\eta}$ and therefore would pick-up a factor of 
$i$ under Wick rotation. The addition of an external symmetry breaking field 
$\lambda X$ to the Lagrangian in Minkowski space becomes then a pure phase 
factor in the path-integral definition of the partition function in Euclidean 
space. But a pure phase factor in the integrand of a partition function 
with positive definite integration measure
can only increase the vacuum energy density and their conclusion was that, 
in such a situation, the mean value of the order parameter should vanish in 
the limit of vanishing symmetry breaking field.

A weak point in this simple and nice argument is the assumption that the 
vacuum energy density (equivalently, the free energy density) is well 
defined when the symmetry breaking external field $\lambda$ is not zero.

We want to show here how Vafa and Witten's argument breaks down if parity 
is spontaneously broken \cite{PARIDAD}. 
In other words, the assumption that the vacuum 
energy density is well defined at non-vanishing $\lambda$ requires the 
previous assumption that parity is not spontaneously broken.

To demonstrate it we will consider the probability distribution function 
(p.d.f.) of any order parameter for parity in vector-like parity-conserving
theories.
The analysis of the p.d.f. of the order parameter in the symmetric model
has been extensively and successful used to investigate spin systems
\cite{BINDER}, spin glasses \cite{GIORGIO} and quantum field theories
with fermionic degrees of freedom \cite{VIC}.

In our approach, in order to work with well defined mathematical 
objects, we will use the lattice regularization scheme and assume 
that the lattice regularized action preserves, as for Kogut-Susskind 
fermions, the positivity of the determinant of the Dirac operator. The 
other essential assumption we use is that the hermitian 
P-non-conserving order parameter is a local operator constructed from 
Bose fields 
and therefore, as any intensive operator, it does not fluctuate in a pure 
vacuum state. This property is equivalent to the statement that all connected 
correlation functions verify cluster property in a pure vacuum state.

The Euclidean path integral formula for the partition function is 

\begin{equation}
\Z = 
\int dA^{a}_{\mu} d\bar\psi d\psi e^{ -\int d^{4}x \left( L(x) + 
i\lambda X(x)\right)}
\label{1}
\end{equation}

\noindent
where following Vafa and Witten we have exhibited the factor of $i$ that 
arises from Wick rotation, i.e. X in (1) is real.

Using the p.d.f. of the order 
parameter $X$, we can write the partition function 
as

\begin{equation}
\Z(\lambda) = \Z(0) \int d \t{X} P( \t{X} ,V) e^{-i \lambda V \t{X}}
\label{2}
\end{equation}

\noindent
where $V$ in (2) is the number of lattice sites, $P(\t{X} ,V)$ is the p.d.f. 
of $X$ at a given lattice volume

\begin{eqnarray}
P(\t{X},V) =\nonumber
\end{eqnarray}
\begin{equation}
\qquad {\int dA^{a}_{\mu} d\bar\psi d\psi e^{-\int d^{4}x L(x)} 
\delta\left(\bar X(A^{a}_{\mu})-\t{X}\right)
\over \int dA^{a}_{\mu} d\bar\psi d\bar\psi e^{-\int d^{4}x L(x)}}
\label{3}
\end{equation}

\noindent
and

$$
\bar X(A^{a}_{\mu}) = \frac{1}{V}\int d^{4}x X(x)\nonumber
$$

Notice that, since the integration measure in (3) is positive or at least 
semi-positive definite, $P(\t{X},V)$ is a true well normalized p.d.f.

Let us assume that parity is spontaneously broken. In the simplest case in 
which there is no extra vacuum degeneracy due to spontaneous breakdown 
of some other symmetry, we will have two vacuum states as corresponds to 
a discrete $Z_2$ symmetry. Since $X$ is an intensive operator, the p.d.f. of 
$X$ will be, in the thermodynamical limit, the sum of two $\delta$ 
distributions:

\begin{equation}
\lim_{V\rightarrow\infty}P(\t{X},V) = 
{1\over2}\delta (\t{X}-a)+{1\over2}\delta (\t{X}+a)
\label{4}
\end{equation}

At any finite volume, $P(\t{X},V)$ will be some symmetric function 
($P(\t{X},V)=P(-\t{X},V)$) developing a two peak structure at 
$\t{X}=\pm a$ and approaching (4) in the infinite volume limit.

Due to the symmetry of $P(\t{X},V)$ we can write the partition function as 

\begin{equation}
\Z(\lambda) = 2\Z(0) Re \int^{\infty}_{0}  P(\t{X},V) e^{-i\lambda V\t{X}}
d\t{X}
\label{5}
\end{equation}

\noindent
and if we pick up a factor of $e^{-i\lambda Va}$
after simple algebra we get:

\begin{eqnarray}
\Z(\lambda)/(2\Z(0)) =\nonumber
\end{eqnarray}
\begin{eqnarray}
\label{7}
& &\cos (\lambda Va)\int^{\infty}_{0}  P(\t{X},V) \cos\left(\lambda 
V(\t{X}-a)\right)d\t{X}\qquad\nonumber \\
&-&\sin (\lambda Va)\int^{\infty}_{0}  P(\t{X},V) \sin\left(\lambda 
V(\t{X}-a)\right) d\t{X}\qquad \nonumber
\end{eqnarray}

The relevant zeroes of the partition function in $\lambda$ can be obtained 
as the solutions of the following equation:

\begin{equation}
\cot (\lambda Va)=
{{\int^{\infty}_{0}  P(\t{X},V) \sin\left(\lambda V(\t{X}-a)\right)
d\t{X}}\over
{\int^{\infty}_{0}  P(\t{X},V) \cos\left(\lambda V(\t{X}-a)\right)
d\t{X}}}
\label{8}
\end{equation}

Let us assume for a while that the denominator in (\ref{8}) is constant at large 
$V$. Since the absolute value of the numerator is bounded by 1, the partition 
function will have an infinite number of zeroes approaching the origin 
($\lambda=0$) with velocity $V$. In such a situation the free energy density 
does not converge in the infinite volume limit.

But this is essentially what happens in the actual case. In fact if we 
consider the integral in (\ref{8}) 

\begin{equation}
f(\lambda V,V)=
\int^{\infty}_{0}  P(\t{X},V) \cos\left(\lambda V(\bar{X}-a)\right)
d\t{X}
\label{9}
\end{equation}

\noindent
as a function of $\lambda V$ and $V$ it is easy to check that the derivative of 
$f(\lambda V,V)$ respect to $\lambda V$ vanishes in the large volume limit 
due to the fact that $P(\t{X},V)$ develops a $\delta(\t{X}-a)$ 
in the infinite
volume limit. At fixed large volumes $V$, 
$f(\lambda V,V)$ as function of $\lambda V$ is 
an almost constant non-vanishing function (it takes the value of $1/2$ at 
$\lambda V=0$). The previous result 
on the zeroes of the partition function in 
$\lambda$ remains therefore unchanged.

To illustrate this result with an example, let us take for $P(\t{X},V)$ a 
double gaussian distribution 

\begin{eqnarray}
P(\t{X},V)=\nonumber
\end{eqnarray}
\begin{equation}
\quad {1\over2}\left({V\over\pi}\right)^{1/2}
\left(e^{-V(\t{X}-a)^{2}} + e^{-V(\t{X}+a)^{2}}\right)
\label{10}
\end{equation}

\noindent
which gives for the partition function 

\begin{equation}
\Z(\lambda) = Z(0) \cos(\lambda Va) e^{-{1\over4}\lambda^{2}V}
\label{11}
\end{equation}

\noindent
and for the mean value of the order parameter 

\begin{equation}
<iX> = {1\over2}\lambda + \tan (\lambda aV) a
\label{12}
\end{equation}

The zeroes structure of the partition function is evident in (\ref{11}) and 
consequently the mean value of the order parameter (\ref{12}) is not defined 
in the thermodynamical limit. Notice also that if $a=0$ (symmetric vacuum), 
the free energy density is well defined at any $\lambda$ and then Vafa 
and Witten's argument applies.

In conclusion we have shown that an essential assumption in the Vafa 
and Witten's theorem on P and CT realization in vector-like 
theories, namely the existence of a free energy density in Euclidean 
space in the 
presence of any external hermitian symmetry breaking source, does not apply 
if the symmetry is spontaneously broken. The assumption that the free energy 
density is well defined requires the previous assumption that the symmetry 
is realized in the vacuum. 

To clarify this point let us discuss a simple model which, as vector-like 
theories, has a positive definite integration measure and, 
after the introduction of an imaginary order parameter, a complex action:
the Ising model in 
presence of an imaginary external magnetic field. This model verifies all 
the conditions of Vafa-Witten theorem. If we assume that the free energy 
density exists, we will conclude that the $Z_2$ symmetry is not spontaneously 
broken. This is obviously wrong in the low temperature phase. The solution 
to this paradox lies in the fact that the free energy density in the low 
temperature phase and for an imaginary magnetic field is not defined (it 
is singular on the imaginary axis of the complex magnetic field plane).
It is true that this model is not a vector-like gauge model but in any case 
verifies all Vafa-Witten conditions, except the existence of the free 
energy density. This example demonstrates that such an assumption is not 
trivial and, what is more relevant, to assume the existence of the free energy 
density is at least at the same level than assume the symmetry be realized 
in the vacuum. 

A possible way to prove Vafa and Witten's 
claim on parity realization in vector-like theories could be to show 
the existence of a Transfer Matrix connecting the Euclidean formulation 
with the Hamiltonian approach
in the presence of any hermitian symmetry breaking field. A weaker sufficient
condition would be the positivity of $Z(\lambda)$ around $\lambda = 0$,
even if, from a mathematical point of view, it is not a necessary one
for the symmetry to be realized.

The proof of these conditions for any symmetry breaking operator seems
very hard, even for the weaker condition. However in the case of the more
standard operator $F\tilde{F}$, associated to the $\theta$-vacuum term,
the reflection positivity of $\Z$ has been shown for the two dimensional
pure gauge model using the lattice  regularization scheme \cite{SEILER}.
Up to our knowledge a generalization of this result to four dimensional
theories and (or) dynamical fermions does not exists.
Only arguments suggesting that a consistent Hamiltonian approach could
be constructed in the four dimensional pure gauge Yang-Mills model
can be found in the literature \cite{ASOREY}.
Summarizing, even if Vafa and Witten's conjecture seems to be plausible,
a theorem on the impossibility to break spontaneously parity in vector-like
theories is still lacking.

\section*{Acknowledgements}

This work has been partially supported by CICYT (Proyecto AEN97-1680). 
A.G. was supported by a Istituto Nazionale di Fisica Nucleare fellowship
at the University of Zaragoza.

\end{document}